\documentclass{aa}

\usepackage{graphicx}
\usepackage{txfonts}
\usepackage{natbib}
\bibpunct{(}{)}{;}{a}{}{,}
\usepackage{stfloats}
\usepackage{amsmath,amssymb}
\usepackage{booktabs}
\usepackage{xcolor,colortbl}
\usepackage{url}
\usepackage[colorlinks=true,urlcolor=blue,citecolor=blue,linkcolor=blue]{hyperref}

\makeatletter
\@ifpackageloaded{lineno}{%
  \nolinenumbers
  \renewcommand\makeLineNumber{}%
  \renewcommand\makeLineNumberLeft{}%
  \renewcommand\makeLineNumberRight{}%
}{}
\makeatother

\begin{document}

\title{The umbrella effect: Magnetic canopy topology modulates spatially averaged chromospheric three-minute power}

\author{
Shahin Jafarzadeh\inst{1,2}\corrauth{shahin.jafarzadeh@qub.ac.uk}
\and David B. Jess\inst{1,3}
\and Marco Stangalini\inst{4}
\and Bernhard Fleck\inst{5}
\and Richard J. Morton\inst{6}
\and Samuel D. T. Grant\inst{1}
\and Peter H. Keys\inst{1}
\and Luc Rouppe van der Voort\inst{2,7}
\and Sven Wedemeyer\inst{2,7}
\and Tiago M. D. Pereira\inst{2,7}
\and Miko{\l}aj Szydlarski\inst{2,7}
\and Elias R. Udn{\ae}s\inst{2,7}
\and Thomas Wiegelmann\inst{8}
\and Markus Roth\inst{9}
}

\institute{
Astrophysics Research Centre, School of Mathematics and Physics, Queen's University Belfast, Belfast BT7 1NN, UK
\and
Rosseland Centre for Solar Physics, University of Oslo, P.O. Box 1029, Blindern, 0315 Oslo, Norway
\and
Department of Physics and Astronomy, California State University Northridge, Northridge, CA 91330, USA
\and
ASI Italian Space Agency, Via del Politecnico snc, I-00133 Rome, Italy
\and
ESA Science and Operations Department, c/o NASA Goddard Space Flight Center, Greenbelt, MD 20771, USA
\and
Department of Mathematics, Physics and Electrical Engineering, Northumbria University, Newcastle upon Tyne NE1 8ST, UK
\and
Institute of Theoretical Astrophysics, University of Oslo, P.O. Box 1029, Blindern, 0315 Oslo, Norway
\and
Max Planck Institute for Solar System Research, Justus-von-Liebig-Weg 3, 37077 G\"ottingen, Germany
\and
Th\"uringer Landessternwarte, Sternwarte 5, 07778 Tautenburg, Germany
}

\date{}

\abstract
{Three-minute ($\sim$5\,mHz) oscillations are a prominent signature of wave propagation in the solar chromosphere, but their measured power can vary substantially between full-disc observations and small, high-resolution fields of view. This variability suggests that the detectability of large-scale chromospheric oscillatory power may depend not only on wave excitation, but also on the magnetic topology through which the waves propagate.}
{We investigate whether chromospheric magnetic canopy topology, in particular its lateral continuity and areal filling, modulates the field-of-view-averaged three-minute power measured in chromospheric diagnostics.}
{We analysed high-resolution H$\alpha$ and Ca~{\sc ii}~8542~\AA\ spectral imaging from the Swedish 1-m Solar Telescope (SST) and compared canopy-poor quiet-Sun scenes with canopy-dominated active-region environments. These observations were complemented by radiative magnetohydrodynamic (MHD) Bifrost simulations representing canopy-poor and canopy-rich atmospheres. We computed field-of-view-averaged power spectra from pixel-wise fixed-wavelength intensities and bisector-derived line-of-sight velocities in the observations, and from synthetic intensities, vertical velocities, and temperature proxies in the simulations. We also quantified the relation between integrated 3--5\,mHz excess power and canopy filling factor in both simulations and observations, using Helioseismic and Magnetic Imager (HMI)-based magnetic-field extrapolations for the latter.}
{In both observations and simulations, canopy-poor scenes show clear enhancements of field-of-view-averaged power in the 3--5\,mHz band, whereas canopy-dominated scenes show strongly reduced measured power in the same frequency range. The trend is present in H$\alpha$ and Ca~{\sc ii}~8542~\AA\ intensity diagnostics and is supported by complementary velocity and temperature proxies. In both the simulations and SST observations, integrated 3--5\,mHz excess power decreases with increasing canopy filling factor.}
{Chromospheric field-of-view-averaged three-minute power is therefore a topology-weighted observable. Dense, laterally continuous magnetic canopies can strongly reduce the measured large-scale three-minute signature without requiring the absence of local wave activity. This effect provides a practical criterion for when chromospheric three-minute power should be recovered in spatially averaged diagnostics and should be considered when interpreting oscillation power as a proxy for atmospheric wave energy in the Sun and, more broadly, in magnetised cool-star atmospheres.}


\authorrunning{Jafarzadeh {et~al.}}
\titlerunning{The umbrella effect}
\maketitle
\nolinenumbers

\section{Introduction}

Oscillations with characteristic periods of approximately three minutes constitute fundamental diagnostic tools for studying wave dynamics and energy propagation in the solar chromosphere. Predominantly excited by upward-propagating acoustic motions driven by turbulent convection beneath the solar surface, these oscillations have been extensively documented across ultraviolet, optical, infrared, and sub-millimetre diagnostics \citep{1991A&A...250..235F, 1991SoPh..134...15R, 1997ApJ...481..500C}. Their ubiquity and coherence have long served as probes of chromospheric heating, atmospheric structuring, and large-scale dynamics \citep{2015SSRv..190..103J}.

High-resolution observations from ground-, balloon-, and space-based facilities have refined our view of wave propagation on small spatial scales \citep{2023LRSP...20....1J}, revealing a broad range of periodicities within localised magnetic structures (sunspots, pores, fibrils, small-scale magnetic elements) and their interactions in the chromosphere \citep{2009Sci...323.1582J, 2014A&A...566A..90M, 2017ApJS..229....7G, 2017ApJS..229....9J, 2018ApJ...869..110S, 2020NatAs...4..220J, 2021RSPTA.37900216S, 2021RSPTA.37900184G, 2022ApJ...930..129B, 2026ApJ..1005L..65J}. Despite this local diversity, field-of-view (FoV)-averaged measures remain valuable for characterising large-scale oscillatory behaviour across different magnetic environments \citep{2009ApJ...702L.168D, 2018AdSpR..61..720G}. Quantitative estimates of the associated wave-energy flux or contribution to chromospheric heating require additional assumptions concerning wave properties and atmospheric response, and are distinct from measurements of oscillatory power itself.

Recent observations with the Atacama Large Millimeter/submillimeter Array (ALMA; \citealt{2009IEEEP..97.1463W}) have revealed substantial variability in the detectability of large-scale three-minute oscillations \citep{2021A&A...656A..68E, 2022FrASS...981205N, 2022A&A...661A..95N}. Clear enhancements occurred only when the observational FoV and its surroundings were magnetically quiet \citep{2021RSPTA.37900174J}, whereas suppression (or absence) of such power correlated with substantial nearby magnetic flux, implicating magnetic topology as a key modulator. Given current uncertainties in precise ALMA formation heights, proposed interpretations included suppression by strong fields, broad formation-height sampling across the FoV, or intrinsically weak temperature fluctuations at these frequencies.

Central to this modulation is the chromospheric magnetic canopy -- an extended, multi-scale magnetic structure rooted in photospheric flux concentrations, whose fields expand with height and become increasingly inclined (often nearly horizontal) at chromospheric heights \citep{1985cdm..proc..175J, 1990A&A...234..519S}, with turnover heights set by the footpoint field strengths \citep{2010ApJ...723L.185W, 2017ApJS..229...11J}. Such canopies facilitate mode conversion, refraction, and reflection where the plasma-$\beta$ approaches unity \citep{2012ApJ...746...68K}. Because of their expansive, inclined geometry over a broad range of spatial scales, canopies may connect to photospheric magnetic concentrations whose one, or even both, footpoints lie outside the observational FoV \citep{2011ApJ...742..119R}, extending their dynamical influence horizontally into the observed region; the resulting FoV-averaged measured three-minute power can therefore depend on their areal coverage. In addition, variations in canopy inclination modulate the effective acoustic cutoff frequency through the ramp effect, enabling or inhibiting upward propagation of three-minute power depending on the local field geometry \citep{1995ApJ...444..879W, 1977A&A....55..239B, 2004Natur.430..536D, 2006ApJ...648L.151J, 2006ApJ...647L..77M}. Recent observations of plage have likewise shown that the estimated chromospheric energy transport and dissipation associated with low-frequency (3--6\,mHz) waves depend on both magnetic-field inclination and strength, with a non-monotonic dependence on inclination \citep{2024ApJ...965..136K}. Such studies quantify local wave-energy transport and dissipation as a function of magnetic-field properties. The present work addresses a complementary question: how the areal filling and lateral continuity of the chromospheric canopy modulate the measured FoV-averaged three-minute power itself.

Three-minute oscillations are well known to occur locally in compact, strongly magnetised structures such as sunspot umbrae and pores, where fields are predominantly vertical and wave guiding is efficient \citep{2019A&A...627A.169F}. The question addressed here is different: we focus on canopy-dominated plage and network environments outside such strong-field cores, where laterally continuous, inclined or horizontal canopies can modulate the measured FoV-averaged three-minute power. While localised oscillation suppression (`magnetic shadows') is also well documented \citep{2001ApJ...561..420M}, a systematic link between canopy topology (spatial extent, inclination, anchoring) and large-scale (FoV-averaged) three-minute power has not been established. Moreover, related localised active-region phenomena such as acoustic halos -- spatially structured enhancements of high-frequency power around strong magnetic concentrations -- are also well established \citep{2011SoPh..268..349S, 2015ApJ...801...27R}, but they address local redistribution of power rather than the FoV-averaged canopy-controlled modulation of the measured three-minute signature that is the focus of the present work. Here we investigate how magnetic topology modulates the measured large-scale three-minute power in the chromosphere by combining high-resolution observations from the Swedish 1-m Solar Telescope (SST; \citealt{2003SPIE.4853..341S}) with advanced Bifrost simulations \citep{2011A&A...531A.154G}. Our multi-height spectral-imaging diagnostics provide strong evidence that magnetic canopies play a critical role in reducing the measured FoV-averaged three-minute power, motivating a re-evaluation of prior interpretations based on lower-resolution data and refining our understanding of chromospheric wave propagation and energy dynamics. We also include ALMA-like synthetic diagnostics in Appendix~\ref{sec:alma_synthetic} to contextualise the observational findings of \citet{2021RSPTA.37900174J}, illustrating how magnetic canopies may similarly influence oscillatory behaviour at millimetre wavelengths. Following earlier usage in \citet{2021RSPTA.37900174J}, and motivated by the canopy-obscuration picture proposed by \citet{2017A&A...598A..89R}, we refer to this canopy-controlled modulation of the measured FoV-averaged three-minute power as the `umbrella effect'. Several processes may contribute (e.g. mode conversion, refraction, and reflection near $\beta\approx1$, inclination-modified cutoff, and phase mixing); our goal is not to isolate a single mechanism but to establish this topology-dependent modulation and a predictive criterion.

These comparisons, together with complementary diagnostics from the simulations (e.g. temperature and line-of-sight (LOS) velocity variations) and LOS velocities from bisector analysis, which offer higher sensitivity to oscillations than fixed-wavelength intensities \citep{1990A&A...228..506D}, provide a multi-faceted view of how magnetic topology modulates wave propagation across atmospheric heights and spectral regimes.
This integrated approach helps separate topology-dependent trends in the measured oscillatory power from diagnostic, radiative-transfer, or projection-related effects that could otherwise complicate the interpretation. Accordingly, we explicitly contrast porous versus laterally continuous canopies (areal filling) to test whether canopy continuity modulates FoV-averaged 3--5\,mHz power in both observations and simulations.

\section{Data and analysis}

\subsection{Observations}
We utilised high-resolution observations acquired with the CRisp Imaging SpectroPolarimeter (CRISP; \citealt{2008ApJ...689L..69S}) mounted on the SST, operating at a spatial sampling of 0.058\,arcsec\,pixel$^{-1}$. Two datasets were analysed: (a) a quiet-Sun time series acquired on 29 May 2020 (07:42--08:34\,UTC), centred at \mbox{($x, y$) = (369, $-$198)\,arcsec} ($\mu=0.90$; $\mu \equiv \cos\theta$, with $\theta$ the heliocentric angle), with an effective full-scan cadence of 23.65\,sec after restoration; and (b) an active region sequence obtained on 8 September 2014 (08:06--09:30\,UTC), centred at \mbox{($x,y$) = ($-$407, $-$323)\,arcsec} ($\mu=0.84$), corresponding to NOAA AR12157, with an effective temporal cadence of 11.57\,sec. These pointings correspond to moderate heliocentric angles, so v$_{\rm LOS}$ is dominated by the vertical component but includes a non-zero horizontal contribution. The quiet-Sun target is an exceptionally magnetically quiet patch embedded within an extended low-activity region,  confirmed by full-disc Solar Dynamics Observatory/Helioseismic and Magnetic Imager (SDO/HMI; \citealt{2012SoPh..275..229S, 2012SoPh..275....3P}) LOS magnetograms. The active-region field samples a strongly magnetised plage (including a few small pores) located between two sunspots within a larger complex active region. For consistency, we restricted analysis to the same central 38$^{\prime\prime}$\,$\times$\,36$^{\prime\prime}$ region in both datasets.

High data quality was achieved owing to excellent seeing and the SST adaptive optics system \citep{2024A&A...685A..32S}. All SST data were processed using the CRISPRED pipeline \citep{2015A&A...573A..40D}, including image restoration using the Multi-Object Multi-Frame Blind Deconvolution (MOMFBD) technique \citep{2005SoPh..228..191V}. This technique mitigates atmospheric seeing effects by estimating the unperturbed solar images from bursts of exposures recorded by multiple synchronised cameras while sampling the spectral sequence.

Each dataset included spectral scans in two chromospheric diagnostics: H$\alpha$\,6563\,\AA\ and Ca\,\textsc{ii}\,8542\,\AA. For H$\alpha$ we analysed the core and five blue-wing offsets at $-$200, $-$400, $-$600, $-$800, and $-$1000\,m\AA; for Ca\,\textsc{ii}\,8542\,\AA, the core and $-$200, $-$400, and $-$600\,m\AA. The selected blue-wing wavelengths have the advantage that photospheric granulation contrast is lower than at the corresponding red-wing locations, improving sensitivity to oscillatory intensity fluctuations. The sampling was identical across datasets and, on average, probed the upper, middle, and lower chromosphere and the mid-to-upper photosphere.

The fixed-wavelength intensity samples were defined relative to the nominal line centres and were applied consistently to both the quiet-Sun and active-region datasets. These intensities should therefore not be interpreted as pure temperature diagnostics: at fixed wavelength they can be modulated by thermodynamic perturbations, opacity changes, and Doppler-induced profile shifts. Their role in this study is comparative. We used the same spectral sampling in the canopy-poor and canopy-dominated scenes, and we interpreted the resulting power spectra together with the bisector-derived velocities, the formation-height estimates, and the synthetic diagnostics.

\subsection{Bisector analysis and LOS velocity diagnostics}
To complement intensity diagnostics, LOS velocity time series were derived from bisector analysis of both H$\alpha$ and Ca\,\textsc{ii}\,8542\,\AA. At each pixel, profiles were interpolated onto a finer wavelength grid and the bisector was computed at fixed fractional intensity levels relative to the local line minimum. For the quiet-Sun dataset, bisectors were extracted at nine levels (10--90\%) in H$\alpha$ and seven levels (10--70\%) in Ca\,\textsc{ii}\,8542\,\AA; for the active-region dataset, 10--80\% in both lines. In the quiet-Sun Ca\,\textsc{ii}\,8542\,\AA\ profiles, levels above about 70\% move into the broad, shallow outer wings beyond the steeper inner-wing portion of the line. Because the local intensity gradient is small there, the inferred wing-crossing wavelengths become highly sensitive to modest pixel-to-pixel variations in profile width, asymmetry, and intensity, producing unstable bisector velocities. We therefore restricted the quiet-Sun Ca\,\textsc{ii}\,8542\,\AA\ analysis to 10--70\%. In the active-region data, the uppermost usable levels were similarly limited by profile asymmetries and noise in strongly magnetised areas. Profiles failing quality checks (e.g. non-monotonic wings, core emission producing multi-valued bisectors) were masked. Bisector levels sample different parts of the line profile and thus, on average, different heights; for H$\alpha$ and Ca\,\textsc{ii}\,8542\,\AA\, the sensitivity tends to concentrate in the photosphere (wings) and upper chromosphere (near-core), with relatively sparse contribution between these extremes (cf. Fig.~\ref{fig:Supp_CFs}). We therefore used the levels as height-ordered proxies rather than fixed geometric heights.

\subsection{Spectral formation heights}
\label{sec:formation_height}
Approximate formation heights for the selected wavelength positions were estimated from line-depression contribution functions (CFs) computed with the one-dimensional version of the RH radiative transfer code \citep{2001ApJ...557..389U, 2015A&A...574A...3P} under non-local thermodynamic equilibrium (non-LTE) conditions using the multilevel accelerated lambda iteration (MALI) scheme \citep{1991A&A...245..171R,1992A&A...262..209R} in the FALC model atmosphere \citep{1993ApJ...406..319F,2009ApJ...707..482F}. Synthetic profiles were convolved with CRISP transmission profiles (full width at half maximum, FWHM, 62\,m\AA\ for H$\alpha$; 107\,m\AA\ for Ca\,\textsc{ii}\,8542\,\AA; \citealt{2015A&A...573A..40D}). Mean formation heights were computed as the CF first moment,
$\langle z\rangle=\frac{\int z\,\mathrm{CF}(z)\,dz}{\int \mathrm{CF}(z)\,dz}\,$.

Across sampled positions the CF widths are typically a few hundred kilometres in the wings and broader near the cores, reflecting finite vertical sensitivity. Tests in FALP (plage) show core-height differences $\lesssim$\,80\,km relative to FALC for both lines, supporting the robustness of the adopted height ordering. For a compact visualisation of the CF computation and representative values, see Fig.~\ref{fig:Supp_CF_cube}.

The full set of CFs and mean formation-height curves for all fixed-wavelength positions and bisector intensity levels is provided in Fig.~\ref{fig:Supp_CFs}. Because line formation depends on the local thermodynamic and magnetic conditions, the effective formation height of a given fixed-wavelength position or bisector level varies in practice across the FoV. Our CF calculations in one-dimensional semi-empirical atmospheres (FALC, with FALP as a plage test) therefore provide spatially averaged height proxies and vertical sensitivity ranges, not per-pixel geometric heights in the observations.
We did not attempt to derive a pixel-by-pixel height distribution over the observed FoV; instead, the intrinsic vertical extent and effective height range of each diagnostic are characterised by the CF widths and the separation between wing and core samples, as illustrated in Fig.~\ref{fig:Supp_CFs}.

\subsection{Numerical simulations}
We complemented the observations with two radiative magnetohydrodynamic (MHD) simulations computed with Bifrost, which solves the full set of MHD equations coupled to non-LTE radiative transfer and field-aligned thermal conduction. The models span the upper convection zone to lower corona with realistic physics and energy balance. The \texttt{en024048} model \citep{2016A&A...585A...4C} represents an enhanced-network configuration with two strong opposite-polarity photospheric flux concentrations that generate pronounced multi-height magnetic canopies. The \texttt{ch024031} model represents a coronal-hole–like configuration with smaller scale, more spatially diffuse magnetic fields that are predominantly vertical with low lateral connectivity, so the chromosphere lacks extended, canopy-like horizontal structuring \citep[for further description of the same simulation see, e.g.][]{2021SoPh..296...84D,2024ApJ...963...10S}. Both domains are 24\,$\times$\,24\,Mm$^2$ horizontally, with $z$ from $\approx-2.4$ to $14.4$\,Mm, with the chromosphere extending up to approximately 2.5\,Mm. We analysed 40 minutes at 10\,s cadence, sufficient to sample three-minute power near 5\,mHz. Because the lower boundary is at $z\approx -2.4$\,Mm, the simulations are not intended to reproduce the Sun's full global $p$-mode cavity and excitation spectrum in detail. Nevertheless, the models do contain strong acoustic and magnetoacoustic fluctuations in the relevant 3--5\,mHz range (e.g. in the coronal-hole case), and our interpretation relies on the relative FoV-averaged chromospheric behaviour of two simulations that share identical numerical and physical treatment but differ in magnetic topology and canopy development.

Synthetic H$\alpha$ and Ca\,\textsc{ii}\,8542\,\AA\ spectra were computed in post-processing. For H$\alpha$, we employed \texttt{SunnyNet}\footnote{\url{https://github.com/tiagopereira/SunnyNet}} framework \citep{2022A&A...658A.182C}, a deep-learning–based tool designed to approximate non-LTE hydrogen populations with high efficiency. \texttt{SunnyNet} utilises convolutional neural networks trained on a few snapshots from the same three-dimensional (3D) MHD simulations run with \texttt{Multi3D} \citep{2009ASPC..415...87L}, capturing realistic chromospheric dynamics and 3D non-LTE radiative transfer effects. It estimates the non-equilibrium level populations of hydrogen at a fraction of the computational cost of full non-LTE solvers. Based on these predicted populations, emergent H$\alpha$ spectra along vertical rays were computed using the \texttt{Muspel.jl}\footnote{\url{https://github.com/tiagopereira/Muspel.jl}} radiative transfer library \citep{2024zndo..10854426P}. For Ca\,\textsc{ii}\,8542\,\AA, we used the RH 1.5D\footnote{\url{https://github.com/ITA-Solar/rh}} code \citep{2015A&A...574A...3P} to solve the non-LTE radiative transfer problem column-by-column to obtain the level populations of Ca\,\textsc{ii}. The final emergent spectra of Ca\,\textsc{ii}\,8542\,\AA\, were calculated with \texttt{Muspel.jl}, which allowed us to include effects of isotopic splitting \citep{2014ApJ...784L..17L}. No instrumental convolution was applied, as our aim is to examine the underlying physical processes rather than replicate a specific observational configuration. We analysed eleven H$\alpha$ blue-wing positions (core to $-$1000\,m\AA\ in 100\,m\AA\ steps) and nine Ca\,\textsc{ii}\,8542\,\AA\ positions (core to $-$800\,m\AA\ in 100\,m\AA\ steps), encompassing (and extending beyond) the observational sampling. 
In addition, temperature and vertical velocity ($v_z$) were extracted at fixed geometric heights every 100\,km from $z{=}100$ to 2500\,km to track vertical evolution of oscillatory power without formation-height ambiguity.

We emphasise that the vertical velocities analysed in the simulations are defined at fixed geometric heights. In the observations, by contrast, LOS velocities were obtained from bisector levels, which sample response-weighted height ranges rather than single geometric layers. Although these diagnostics therefore differ in their effective formation heights, our comparison focuses on relative trends in FoV-averaged 3--5\,mHz power across canopy topologies, and our main conclusions do not rely on an exact one-to-one correspondence of formation heights between simulations and observations.

\subsection{Magnetic field diagnostics}
The magnetic environment of the observations was characterised using SDO/HMI vector data. Full-Stokes parameters were inverted with the Very Fast Inversion of the Stokes Vector (VFISV) \citep{2011SoPh..273..267B} to obtain photospheric vector fields. To reconstruct topology over domains considerably larger than the SST FoV, we applied a potential-field extrapolation for the quiet-Sun dataset and a non-linear force-free field (NLFFF) extrapolation for the active-region dataset \citep{2005A&A...433..701W, 2018ApJ...866..130Z}, chosen to match the magnetic regime (strongly non-potential active region; near-potential quiet Sun). This also enabled mapping of extended canopy structures rooted outside the observed areas but possibly inclined into them at chromospheric heights. We did not rely on chromospheric spectropolarimetric inference here, as quiet-Sun chromospheric Stokes signals typically have insufficient signal-to-noise for robust inclination constraints; moreover, our goal is a consistent multi-height canopy topology over a domain larger than the SST FoV, for which HMI-based extrapolations provide a uniform photospheric boundary condition. In addition, the orientation of fibrillar structures in H$\alpha$ and Ca\,\textsc{ii}\,8542\,\AA\ intensity images can be used as a qualitative proxy for field direction at higher layers, providing morphological constraints complementary to the extrapolations. For the simulations, the full 3D magnetic field is known at all heights, allowing a direct reconstruction of canopy geometry and its evolution.

\subsection{Data analysis}
At each spatial pixel, we extracted time series of (i) intensity (observations and synthetic spectra), (ii) bisector-derived LOS velocity (observations), and (iii) temperature and vertical velocity from simulations at fixed geometric heights (with $v_z$ serving as a disc-centre LOS proxy). For the off-centre SST observations, LOS velocities contain both vertical and horizontal components. We therefore used the velocity spectra as complementary diagnostics and based the primary topology comparison on intensity spectra, which are not subject to the same direct vector-projection ambiguity as velocities, although fixed-wavelength intensities can still be influenced by Doppler-induced profile shifts and thermodynamic changes of the line profile. These observables predominantly trace compressive signatures; thus, reduced power in a given diagnostic can reflect a reduced measured response (e.g. via mode conversion, phase mixing, or changes in diagnostic sensitivity) rather than a complete absence of wave energy. Time series were linearly detrended and apodised with a Tukey window (shape parameter 0.1), then transformed using a fast Fourier transform (FFT) to obtain power spectra. Powers were normalised by the frequency resolution to yield power per mHz (power spectral density). FoV-averaged (large-scale) spectra were computed as the arithmetic mean of pixel-wise power spectra (not time-domain averages), so the presence or absence of a 3--5\,mHz peak is not an artefact of time-domain phase cancellation. Here and throughout, suppression refers operationally to a reduction of the measured FoV-averaged 3--5\,mHz power in the analysed diagnostics. We use the term three-minute more generally to describe the chromospheric oscillatory enhancement centred around $\sim$5\,mHz, whereas references to the 3--5\,mHz band denote the specific frequency interval adopted for the quantitative band-power and excess-power measurements. It does not imply, by itself, that local wave activity or wave energy is absent; reduced measured power may instead reflect changes in propagation, mode partition, dissipation or redistribution of wave power, or diagnostic sensitivity. As cross-checks, Morlet wavelet transforms (retaining only power unaffected by the cone of influence, and significant at the 95\% level) and empirical mode decomposition and Hilbert–Huang analysis yielded trends consistent with the FFT results. All analysis steps used the open-source WaLSAtools\footnote{\url{https://github.com/WaLSAteam/WaLSAtools}} package \citep{2025NRvMP...5...21J, walsatools..2025...17569951}.

Because the FoV-averaged spectra combine a large number of pixel-wise power spectra, averaging reduces the statistical fluctuations associated with uncorrelated noise, while any mean noise-power contribution remains as a background component. Our results therefore rely primarily on relative differences in the 3--5\,mHz spectral behaviour between canopy topologies rather than on absolute noise-calibrated power levels. In the simulations, which contain no observational photon or detector noise, the corresponding trends provide an independent comparison free from these observational noise sources.

\section{Results}

We first compare the measured FoV-averaged three-minute power in canopy-poor and canopy-dominated SST observations, and then test the same topology dependence in Bifrost simulations. The primary comparison is based on H$\alpha$ and Ca~{\sc ii}~8542\,\AA\ intensity spectra, with bisector-derived LOS velocities and simulation temperature and velocity proxies used as complementary diagnostics. Representative spatial maps of the pixel-wise 3--5\,mHz band power are shown in Fig.~\ref{fig:Supp_bandpower}, and the corresponding relations between integrated 3--5\,mHz excess power and canopy filling factor in the simulations and observations are quantified in Figs.~\ref{fig:supp_int_excess_ff} and \ref{fig:supp_int_excess_ff_obs}, respectively.

\subsection{Quiet-Sun observations}

\begin{figure*}[!]
\centering
\includegraphics[width=0.965\textwidth]{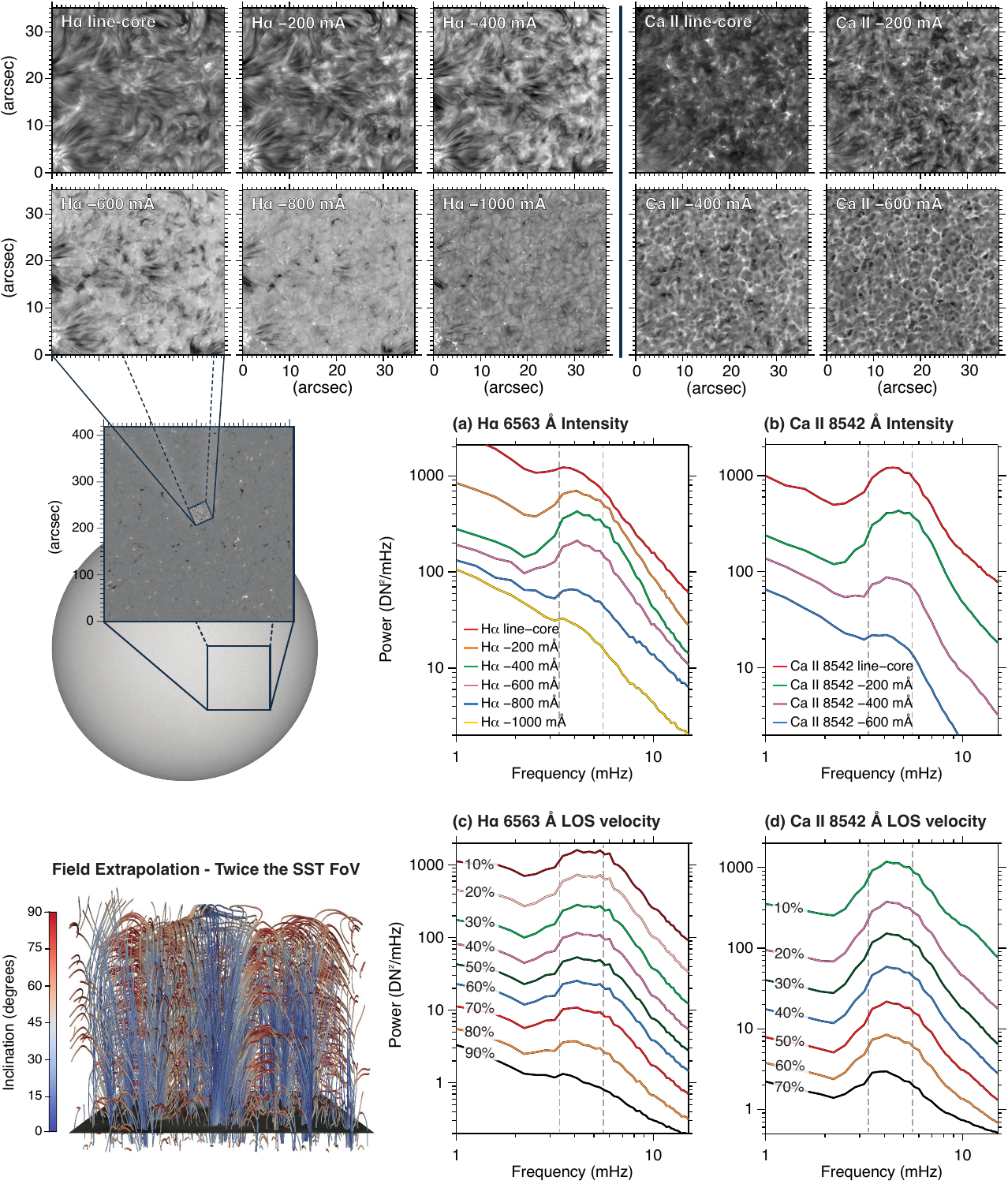}
\caption{FoV-averaged oscillatory power in a quiet-Sun region (SST).
Top: SST/CRISP intensity maps in H$\alpha$ (core to $-1000$\,m\AA) and Ca\,\textsc{ii}\,8542\,\AA\ (core to $-600$\,m\AA), sampling different heights.
Middle: Left: SST FoV (38$^{\prime\prime}$\,$\times$\,36$^{\prime\prime}$) on SDO/HMI magnetogram and full-disc continuum.
(a,b): FoV-averaged intensity power spectra (log--log) for H$\alpha$ and Ca\,\textsc{ii}\,8542\,\AA\ at all sampled offsets.
Bottom: Left: Side-on 3D extrapolated magnetic field above the photospheric image plane ($z=0$). The horizontal domain spans 76$^{\prime\prime}$\,$\times$\,72$^{\prime\prime}$, corresponding to twice the SST FoV in each direction, and the extrapolation shown extends vertically to $z\simeq2.8$\,Mm. Field lines are colour-coded by inclination (blue: near-vertical; red: near-horizontal), revealing a fragmented topology with only short, locally rooted canopies. The vertical dimension is visually exaggerated for clarity and is not plotted to the same scale as the horizontal dimensions.
(c,d): FoV-averaged bisector-derived LOS-velocity power spectra for H$\alpha$ and Ca\,\textsc{ii}\,8542\,\AA\ (90\%--10\% for H$\alpha$ and 70\%--10\% for Ca~\textsc{ii}~8542\,\AA; bottom to top). Strong three-minute ($\sim$5\,mHz) enhancements peak in the mid-to-upper chromosphere. For reference, vertical dashed lines at 3.3 and 5.5\,mHz (approximate five- and three-minute periods) are overplotted. All power spectra in panels (a)--(d) are vertically displaced for clarity by 0.3\,dex between successive curves (a factor of $10^{0.3}\simeq2$); the lowest curve in each panel is unshifted, and the absolute $y$-axis power scale refers to that curve.}
\label{fig:QS_SST}
\end{figure*}

Fig.~\ref{fig:QS_SST} summarises the quiet-Sun dataset. The top panels show six H$\alpha$ snapshots (core and $-200$, $-400$, $-600$, $-800$, $-1000$\,m\AA), and four Ca\,\textsc{ii}\,8542\,\AA\ snapshots (core and $-200$, $-400$, $-600$\,m\AA). These positions sample progressively deeper atmospheric layers away from the line core.

In both diagnostics, line-core images more clearly exhibit fibrillar structures -- especially in H$\alpha$ -- that trace the magnetic topology \citep{2011A&A...527L...8D,2013ApJ...776...56R}. The fibrils are sparse and do not densely fill the region, consistent with a weak, porous canopy. With increasing blue-wing offset, fibrils fade and are nearly absent in the far wing, reflecting the lower formation heights where magnetic structuring is less apparent in intensity. Thus, the line cores provide the clearest chromospheric manifestation of the underlying, fragmented topology.

The middle-left panels outline the SST FoV on a larger SDO/HMI magnetogram within a full-disc continuum image, confirming the magnetic quietness of the surroundings. The bottom-left panel shows a side-on 3D extrapolation from the photosphere to $\sim$3\,Mm, with field lines colour-coded by inclination (blue: near-vertical; red: near-horizontal). The topology is fragmented, with intermittent multi-height canopies rooted in small magnetic elements, but no laterally continuous, dense canopy over the SST FoV. The plotted domain is twice the SST FoV to capture connectivity beyond the scene.

Panels~(a,b) display FoV-averaged power spectra from intensity time series in H$\alpha$ and Ca\,\textsc{ii}\,8542\,\AA\, (log–log scale; vertically offset). Both diagnostics show a strong enhancement near 5\,mHz at mid-chromospheric positions. In H$\alpha$, power grows from a mild $\sim$3.5\,mHz enhancement at $-1000$\,m\AA\ to a peak near 4\,mHz at $-800$\,m\AA, maximises around 4.5--5\,mHz at $-400$\,m\AA, and weakens slightly at $-200$\,m\AA; the line core nearly resembles $-800$\,m\AA, suggesting some suppression/redirection at upper chromospheric heights. Ca\,\textsc{ii}\,8542\,\AA\, shows a similar trend: modest $\sim$4\,mHz enhancement at $-600$\,m\AA, increasing at $-400$\,m\AA\ and peaking near 5\,mHz at $-200$\,m\AA; the core resembles $-400$\,m\AA.

Panels~(c,d) show FoV-averaged power spectra from bisector-derived LOS velocities. For H$\alpha$ (nine levels, 90\% to 10\% of line depth), the peak shifts from $\sim$3.5\,mHz (90\%) to $\sim$4\,mHz (80\%) and to 4.5--5\,mHz at higher layers (30--10\%), with the strongest enhancement near 20\%. Ca\,\textsc{ii}\,8542\,\AA\, (seven levels, 70--10\%) behaves similarly, again peaking near 5\,mHz at mid-to-upper chromospheric levels (strongest around 20\%). Together, the intensity and Doppler diagnostics are consistent with height-dependent propagation of large-scale three-minute oscillatory power in the absence of a dense overlying canopy.

\subsection{Active-region observations}

\begin{figure*}[!]
\centering
\includegraphics[width=0.97\textwidth]{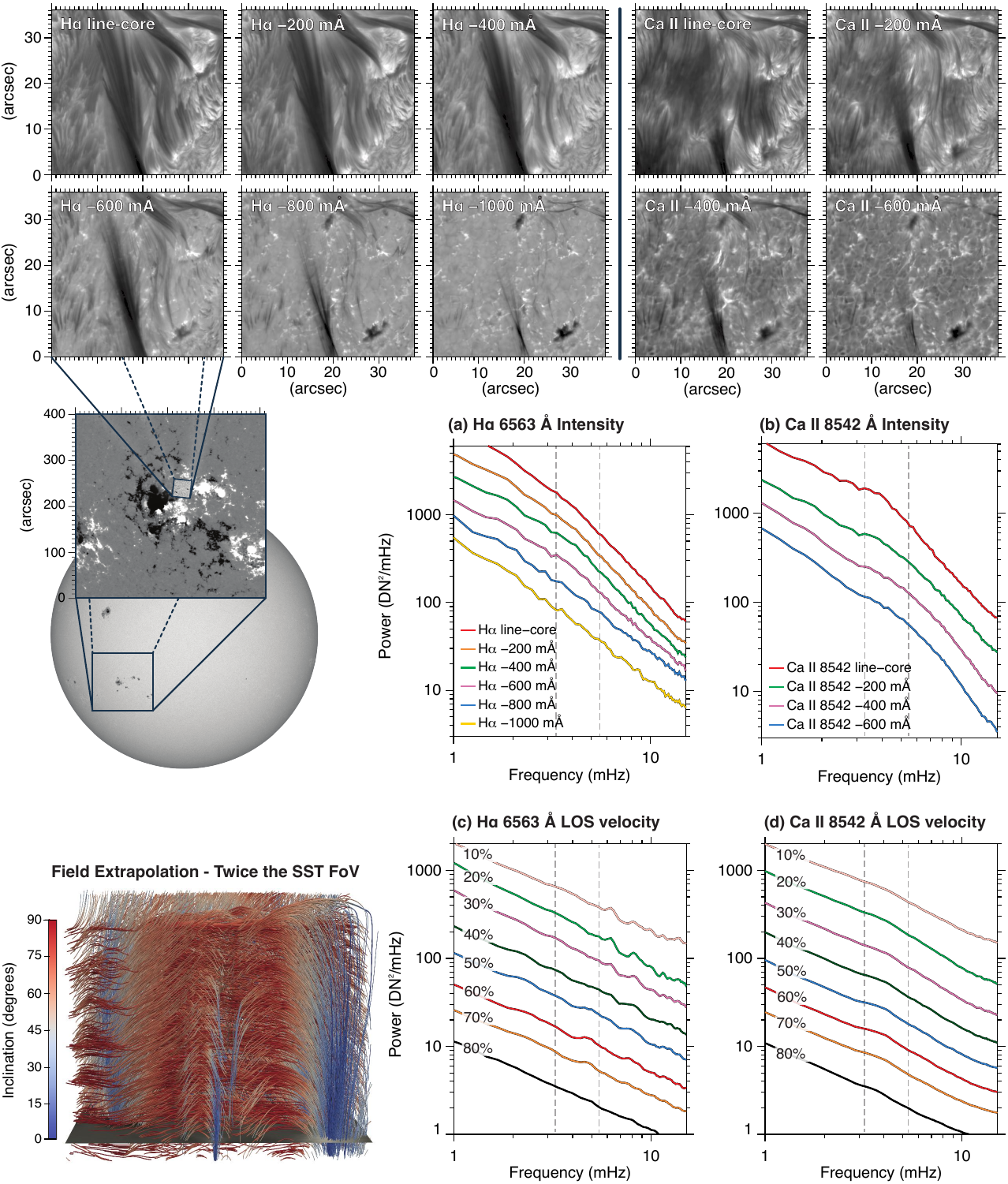}
\caption{Suppression of FoV-averaged oscillatory power in an active region.
Top: SST/CRISP intensity maps in H$\alpha$ and Ca\,\textsc{ii}\,8542\,\AA\ from core to blue wing.
Middle: Left: SST field of view (38$^{\prime\prime}$\,$\times$\,36$^{\prime\prime}$) on SDO/HMI magnetogram and full-disc continuum, showing nearby strong flux.
(a,b): FoV-averaged intensity power spectra (log--log); no significant enhancement in the 3--5\,mHz band.
Bottom: Left: Side-on 3D extrapolated magnetic field above the photospheric image plane ($z=0$). The horizontal domain spans 76$^{\prime\prime}$\,$\times$\,72$^{\prime\prime}$, corresponding to twice the SST FoV in each direction, and the extrapolation shown extends vertically to $z\simeq2.8$\,Mm. Field lines are colour-coded by inclination (blue: near-vertical; red: near-horizontal), revealing dense, multi-height, laterally continuous canopies over the SST FoV. The vertical dimension is visually exaggerated for clarity and is not plotted to the same scale as the horizontal dimensions.
(c,d): Bisector-derived LOS-velocity power spectra (80\%--10\% levels); no prominent peaks at any height, indicating suppression of large-scale three-minute power. Vertical dashed lines at 3.3 and 5.5\,mHz are shown as visual guides (approximate five- and three-minute reference periods). All power spectra in panels (a)--(d) are vertically displaced for clarity by 0.3\,dex between successive curves (a factor of $10^{0.3}\simeq2$); the lowest curve in each panel is unshifted, and the absolute $y$-axis power scale refers to that curve.}
\label{fig:AR_SST}
\end{figure*}

Fig.~\ref{fig:AR_SST} (same layout as Fig.~\ref{fig:QS_SST}) shows a striking contrast. Line-core images of both H$\alpha$ and Ca\,\textsc{ii}\,8542\,\AA\, are dominated by dense, long fibrils with high areal filling -- a laterally continuous canopy -- particularly in H$\alpha$, with filaments spanning the FoV. These features persist into nearby wing positions, evidencing strong, extended canopy fields; far-wing images show fewer such structures, consistent with lower heights. The continuity of line-core fibrils is a strong proxy for dense canopy topology characterising this active region.

The middle-left panel outlines the SST FoV on an SDO/HMI magnetogram. The bottom-left panel shows a side-on 3D extrapolation from the photosphere to $\sim$3\,Mm over a domain twice the SST FoV: numerous field lines extend horizontally, forming multi-height canopies with high lateral continuity -- densest over the SST FoV (the central part of the domain) where long, nearly horizontal bundles create an overlying canopy at several heights.

Panels~(a,b) show FoV-averaged intensity power spectra in H$\alpha$ and Ca\,\textsc{ii}\,8542\,\AA. Unlike the quiet Sun, there is no pronounced $\sim$5\,mHz enhancement; spectra are largely flat, with only weak knees or inflections near 3.5--4\,mHz at mid-chromospheric positions (more evident in Ca\,\textsc{ii}). Panels~(c,d) show bisector-derived LOS velocities (80--10\%): again no clear 3--5\,mHz enhancement in either line. Thus, in this canopy-dominated FoV, large-scale three-minute signatures are not detected in intensity or Doppler diagnostics.

While localised transient dynamics can affect power locally, the suppression reported here is systematic across two independent diagnostics and is reproduced in the canopy-dominated simulation (see below), supporting a topology-driven interpretation. We emphasise that the analysed active-region subfield samples a canopy-dominated plage and network environment outside the compact strong-field cores and is not intended to provide a statistical representation of all active-region environments. Strong local three-minute power may still occur in or around compact magnetic concentrations, pores, or sunspot umbrae where the field is predominantly vertical. Our conclusion from this example is therefore not that active regions generally lack three-minute oscillations, but that a FoV dominated by laterally extended inclined canopies can show strongly reduced measured FoV-averaged three-minute power.

\subsection{Quiet-Sun (coronal hole) simulation}

\begin{figure*}[h]
\centering
\includegraphics[width=1.0\textwidth]{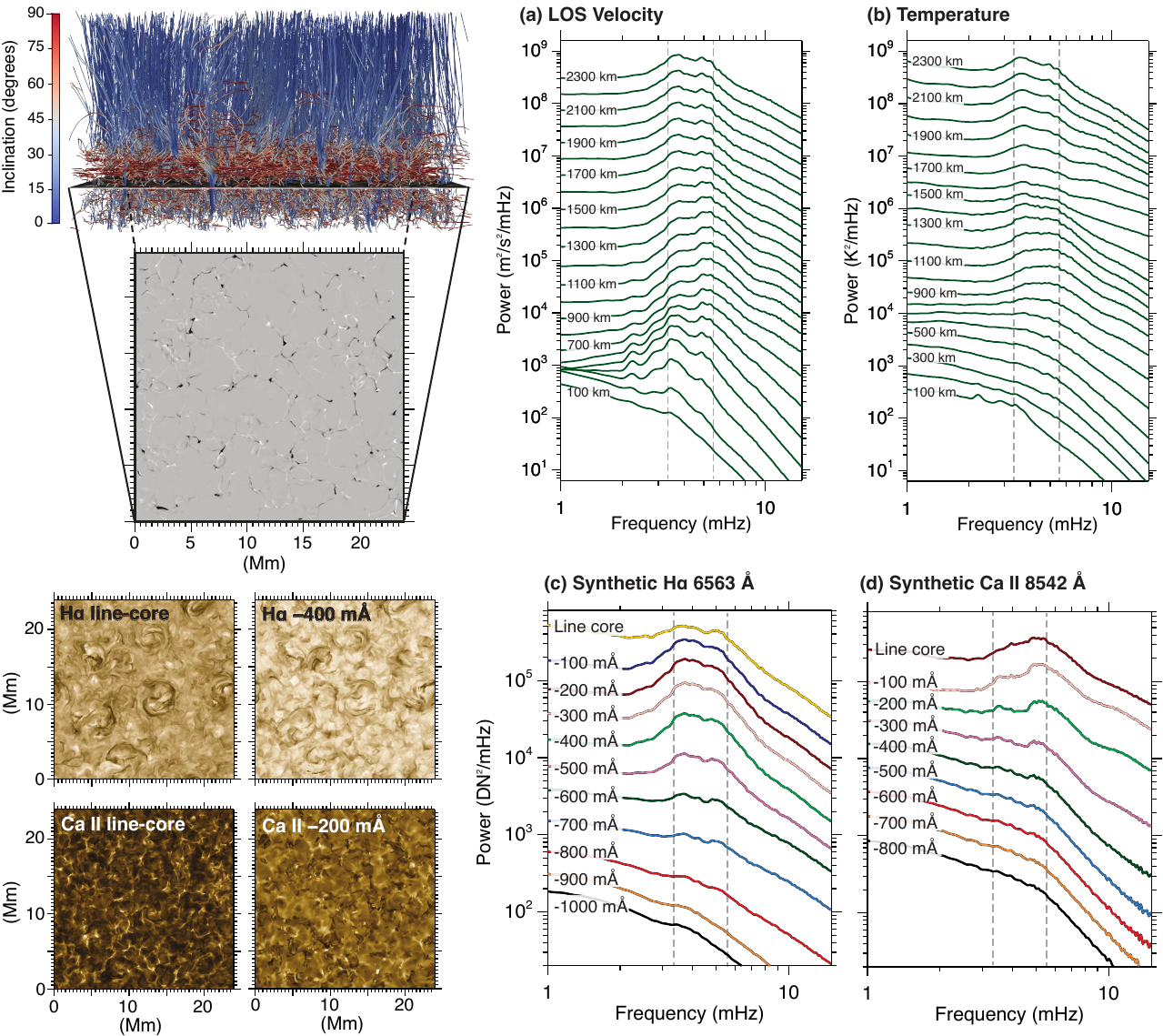}
\caption{Large-scale oscillation signatures in a quiet-Sun (coronal-hole) simulation.
Top left: Photospheric $B_z$ and side-on field topology (inclination-coloured), showing absence of dense chromospheric canopies.
Top middle and right: Height-resolved FoV-averaged power spectra of LOS velocity and temperature (100--2300\,km), showing the development of a pronounced $\sim$3--5\,mHz chromospheric enhancement and its evolution with height.
Bottom middle and right: Power spectra from synthetic H$\alpha$ and Ca\,\textsc{ii}\,8542\,\AA\ intensities across eleven and nine offsets (vertically offset); strong $\sim$5\,mHz enhancements in the mid-chromosphere, weakening towards lower and upper layers.
Bottom left: Representative synthetic images at selected offsets. Vertical dashed lines indicate 3.3 and 5.5\,mHz as approximate five- and three-minute reference frequencies. For clarity, successive power spectra in panels (a)--(d) are offset vertically by 0.3\,dex (a factor of $10^{0.3}\simeq2$). The lowest curve in each panel is shown without an offset and defines the absolute $y$-axis power scale.}
\label{fig:QS_sim}
\end{figure*}

Fig.~\ref{fig:QS_sim} provides a complementary canopy-poor reference simulation rather than a one-to-one reproduction of the observed quiet-Sun magnetic topology. The top-left panels show the lower-photospheric magnetic map and a 3D field topology (inclination-coloured). Although the detailed field morphology differs from the extrapolated quiet-Sun observation in Fig.~\ref{fig:QS_SST}, the key common property is the absence of a dense, laterally continuous chromospheric canopy over the analysed FoV. In this sense, the simulation tests whether a canopy-poor atmosphere produces the same type of measured FoV-averaged three-minute enhancement as found in the observations.

Height-resolved power spectra of LOS velocity and temperature (100--2300\,km; panels a,b) show a strong $\sim$5\,mHz enhancement in the chromosphere, especially between 1100--1300\,km, showing the height-dependent development of large-scale three-minute oscillatory power. In the photosphere ($\lesssim$700\,km), power is weaker and peaks near 3--4\,mHz, shifting upward in frequency with height before gradually declining towards the transition region.

Synthetic spectral diagnostics mirror this: H$\alpha$ (panel c) shows little enhancement at $-1000$ to $-800$\,m\AA, followed by a steady increase in power enhancement and a shift in peak frequency with height (between 4.5 and 5\,mHz), maximising around $-400$ to $-300$\,m\AA, and weakening at the core. Ca\,\textsc{ii}\,8542\,\AA\, behaves similarly: the first 4--5 wing positions show no clear peak, then a knee shifts towards 5\,mHz; from $-300$\,m\AA\ upward a clear $\sim$5\,mHz peak emerges, maximising near $-100$\,m\AA\ and persisting at the core.

\subsection{Active-region (enhanced network) simulation}

\begin{figure*}[h]
\centering
\includegraphics[width=1.0\textwidth]{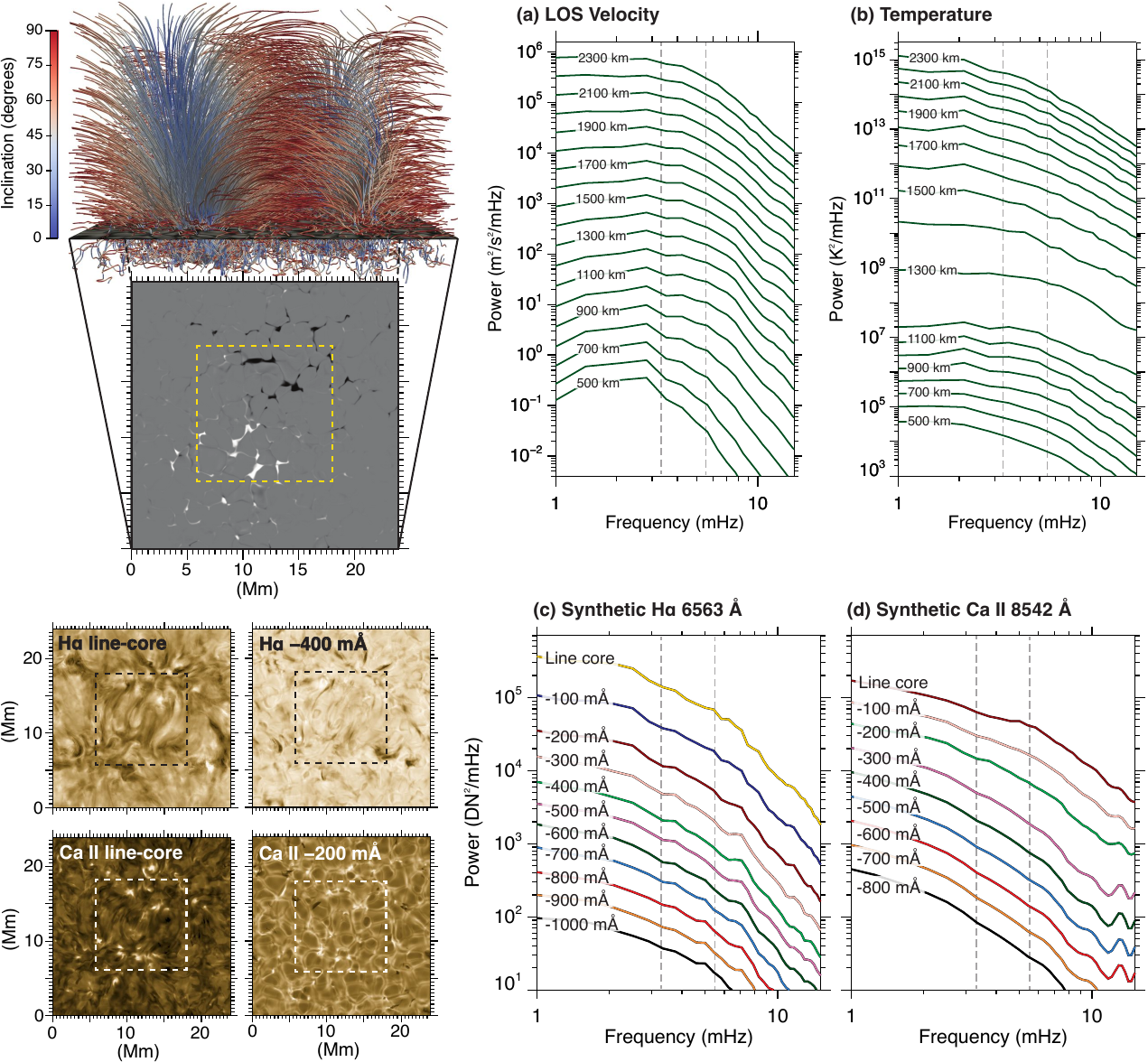}
\caption{Multi-height canopy–driven suppression of large-scale oscillations in an enhanced-network simulation.
Top left: Photospheric $B_z$ and field topology showing a well-developed, multi-height canopy formed by opposite-polarity patches; the canopy-dominated analysis region is marked.
 Top middle and right: Height-resolved FoV-averaged power spectra of LOS velocity and temperature (500--2300\,km), showing smooth spectral profiles without a pronounced 3--5\,mHz power enhancement.
Bottom middle and right: Power spectra from synthetic H$\alpha$ and Ca\,\textsc{ii}\,8542\,\AA\, across sampled offsets (vertically offset); no significant 3--5\,mHz peaks.
Bottom left: Representative synthetic images at selected offsets. Vertical dashed lines mark 3.3 and 5.5\,mHz as approximate five- and three-minute reference periods. The spectra in panels (a)--(d) are vertically staggered by 0.3\,dex between successive curves for visual clarity (equivalent to a factor of $10^{0.3}\simeq2$). In each panel, the bottom curve is unshifted and sets the absolute $y$-axis power scale.}
\label{fig:AR_sim}
\end{figure*}

Fig.~\ref{fig:AR_sim} shows dense, multi-height canopies from the photosphere into the chromosphere, rooted in strong opposite-polarity flux. The 3D field rendering displays horizontally extended structures with high lateral continuity across the central analysis region. We restricted analysis to this canopy-dominated mask (excluding quieter edges; cf. \citealt{2021A&A...656A..68E}).

Height-resolved LOS velocity and temperature spectra (500--2300\,km) show no clear $\sim$5\,mHz enhancement. Instead, spectra are smooth (log--log), with subtle turning points at lower frequencies and no pronounced three-minute power excess. A weak knee around $\sim$2.8\,mHz is visible in some of the $v_z$ spectra. The enhanced-network simulation is known to contain comparatively strong global oscillations with periods of approximately 350--500\,s, associated with the finite simulation domain and its lower-boundary treatment \citep{2016A&A...585A...4C}; we therefore do not interpret this low-frequency feature as a chromospheric three-minute power enhancement.

Synthetic H$\alpha$ and Ca\,\textsc{ii}\,8542\,\AA\, power spectra across all sampled offsets are largely featureless near 5\,mHz. In some Ca\,\textsc{ii} spectra at lower heights we note weak, narrow features above 10\,mHz; they lack consistent counterparts across diagnostics and lie outside our focus on broad 3--5\,mHz power, so we do not interpret them further.  Overall, the canopy-rich atmosphere is associated with a strong reduction of the measured large-scale three-minute power in the chromospheric diagnostics.

\subsection{Integrated 3--5\,mHz power excess versus canopy filling factor}
\label{sec:supp_int_excess_ff}
To provide a quantitative test of the umbrella effect criterion, we measured how the 3--5\,mHz power enhancement in chromospheric intensity at fixed wavelength offsets depends on the canopy filling factor within the same field of view. For both the canopy-poor quiet-Sun simulation and the canopy-dominated enhanced-network simulation, we used the representative chromospheric sampling positions shown in Fig.~\ref{fig:Supp_bandpower}: Ca~{\sc ii}~8542\,\AA\ intensity at $-200$\,m\AA\ and H$\alpha$ intensity at $-400$\,m\AA. The same wavelength offsets were used for the corresponding SST observations. We then computed an `excess power' metric ($EP$) following \citet{2017ApJ...847....5K}: at each pixel we fitted a power-law background to the power spectrum ($P$) outside the frequency interval used to measure the excess and then evaluated the bandwidth-normalised integrated excess within 3--5\,mHz,
\begin{equation}
EP \;=\; \frac{1}{5\,\mathrm{mHz}-3\,\mathrm{mHz}}
\int_{3\,\mathrm{mHz}}^{5\,\mathrm{mHz}} \max\!\left(\frac{P(f)}{P_{\rm fit}(f)}-1,\,0\right)\,df \, .
\end{equation}

We quantified canopy coverage using the canopy filling factor, \(ff\), defined as the fraction of pixels in a given spatial window that satisfy a simple canopy criterion based on the magnetic-field strength ($B$) and inclination angle ($\gamma$) at chromospheric heights:
\begin{equation}
ff \;=\; \frac{N\!\left(\,B \ge B_{\min}\;\wedge\;\gamma \ge \gamma_{\min}\,\right)}{N_{\rm valid}} \, ,
\end{equation}
where \(N_{\rm valid}\) counts pixels with finite values, and we adopted representative thresholds \(B_{\min}=5\,\mathrm{G}\) and \(\gamma_{\min}=50^{\circ}\), evaluated at \(\sim\)1.5\,Mm in the simulations; the corresponding observational implementation at 1.44\,Mm is described below. Here $\gamma$ is measured relative to the local vertical, such that $\gamma=0^{\circ}$ denotes a vertical field and $\gamma=90^{\circ}$ a horizontal field. The dimensionless pixel-wise \(EP\) values were then averaged within sliding spatial windows (fixed physical window size and step), and compared with the corresponding window-averaged canopy filling factor \(ff\). 

For the enhanced-network simulation, we followed the same field-of-view selection adopted throughout the paper and restricted the analysis to the central 50\% of the synthetic domain (in both spatial directions), because the outer parts include comparatively canopy-poor regions that would otherwise dilute the canopy-rich regime we aimed to characterise and bias the inferred $EP$--$ff$ relationship. For the simulations, varying the adopted thresholds and the sampling height within reasonable ranges did not change the qualitative behaviour reported below, but only shifted the numerical values of \(ff\) by modest amounts.

\begin{figure*}[h]
\centering
\includegraphics[width=1.0\textwidth]{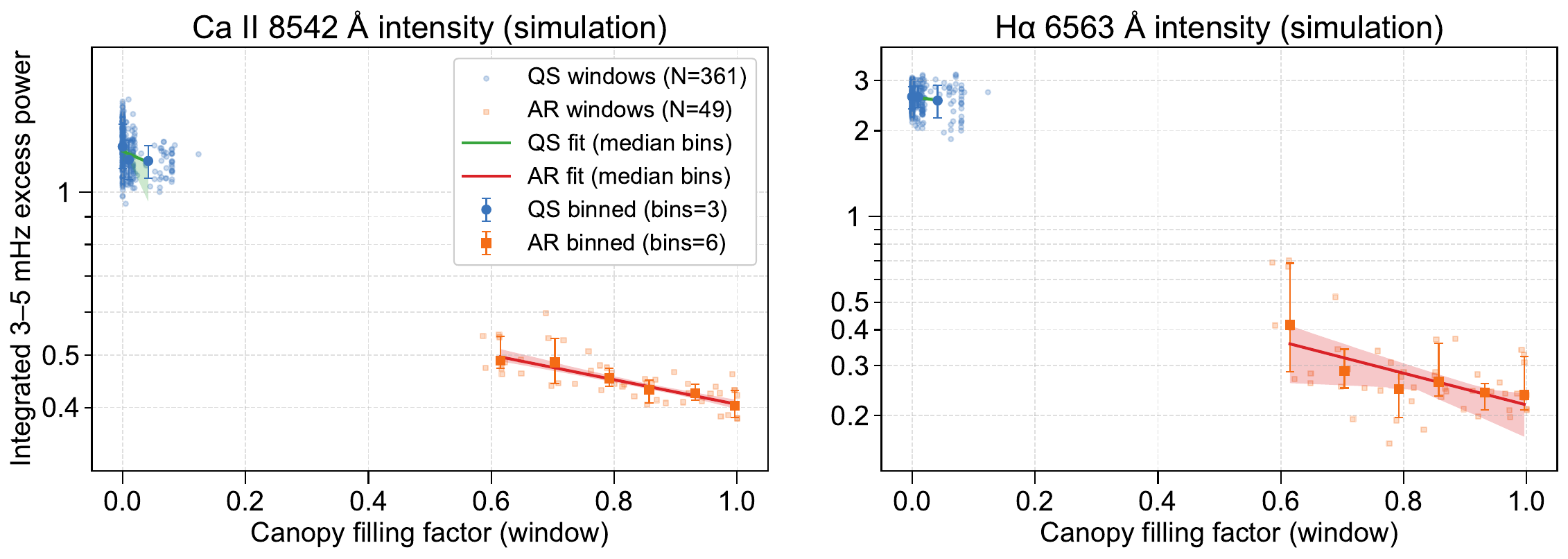}
\caption{Integrated 3--5\,mHz power excess versus canopy filling factor in simulations.
Scatter plots show the integrated excess power in the 3--5\,mHz band, measured from synthetic intensity time series in Ca~{\sc ii}~8542\,\AA\ at $-200$\,m\AA\ (left) and H$\alpha$ at $-400$\,m\AA\ (right), versus the chromospheric canopy filling factor $ff$ computed from magnetic-field strength and inclination angle thresholds. Points correspond to sliding spatial windows across each simulation field of view. Binned medians and fitted log-space trends illustrate the systematic decrease of integrated excess power with canopy filling factor in both diagnostics.}
\label{fig:supp_int_excess_ff}
\end{figure*}

Fig.~\ref{fig:supp_int_excess_ff} shows the resulting relationship between integrated 3--5\,mHz excess power and canopy filling factor for both lines in simulations. In all cases, the integrated excess power decreases with increasing canopy filling factor, consistent with the umbrella effect expectation that laterally continuous inclined canopies reduce the measured FoV-averaged three-minute signature.

For the adopted windowing in the simulations, we obtain 361 windows in the canopy-poor coronal-hole simulation and 49 in the enhanced-network simulation for each line. These counts follow directly from the chosen physical window size and step, applied over the full coronal-hole simulation domain and over the central-cropped (50\%) enhanced-network domain; the smaller analysed domain in the latter therefore yields fewer valid window placements. In Ca~{\sc ii}~8542\,\AA, the canopy-poor simulation windows cluster at very low filling factor (median \(ff \approx 0.00\), spanning 0.00--0.12) and show substantially larger integrated excess (median \(\approx 1.19\)) than the enhanced-network windows (median \(ff \approx 0.83\), spanning 0.59--1.00; median excess \(\approx 0.44\)), corresponding to a median canopy-poor to enhanced-network ratio of \(\approx 2.7\) (0.43\,dex). In H$\alpha$, the separation is even stronger: the canopy-poor simulation median integrated excess (\(\approx 2.62\)) exceeds the enhanced-network median (\(\approx 0.26\)) by a factor of \(\approx 10\) (1.00\,dex). Linear trends fitted in log-space are negative for both simulations and both diagnostics, showing an overall decrease of 3--5\,mHz excess power with increasing canopy filling factor.

For the observations, the canopy filling factor was estimated from the HMI-based magnetic-field extrapolations. The photospheric magnetic maps were co-registered with the SST FoVs using the large-scale magnetic morphology, and the same geometrical transformations were then applied to the extrapolated $B_x$, $B_y$, and $B_z$ components at 1.44\,Mm. The high-resolution SST $EP$ maps were spatially averaged onto the native extrapolation grid (360\,km sampling), rather than upsampling the lower-resolution magnetic field. For the observational analysis, we used approximately 8\,Mm spatial windows stepped by approximately 2\,Mm (22 and 6 extrapolation pixels, corresponding to 7.92 and 2.16\,Mm, respectively), and applied the same nominal $B_{\min}=5$\,G and $\gamma_{\min}=50^{\circ}$ criteria as for the simulations. Because the observational magnetic field is inferred from HMI-based extrapolations rather than known directly, we additionally verified that the qualitative $EP$--$ff$ relationship is insensitive to reasonable variations of the adopted field-strength threshold.

Because adjacent sliding windows overlap in both the simulation and observational analyses, individual points are not statistically independent; the windowed analysis was therefore used to characterise the spatial relationship between $EP$ and $ff$, rather than to assign formal significance from the number of windows alone.

\begin{figure*}[h]
\centering
\includegraphics[width=1.0\textwidth]{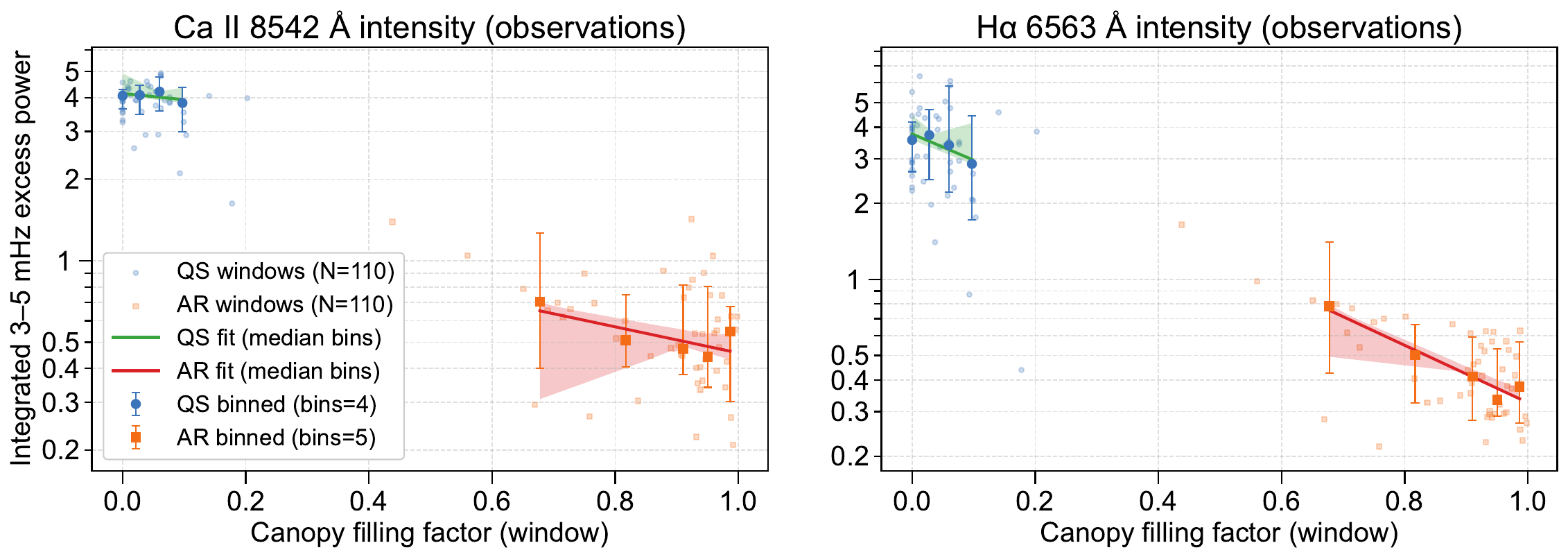}
\caption{Integrated 3--5\,mHz power excess versus canopy filling factor for the SST observations. The analysis is analogous to Fig.~\ref{fig:supp_int_excess_ff}: the integrated excess power in the 3--5\,mHz band is measured from Ca~{\sc ii}~8542\,\AA\ intensity at $-200$\,m\AA\ (left) and H$\alpha$ intensity at $-400$\,m\AA\ (right), while the canopy filling factor is derived from the co-registered HMI-based magnetic-field extrapolations at 1.44\,Mm using the same $B\ge5$\,G and $\gamma\ge50^{\circ}$ criteria. Points correspond to sliding spatial windows across the observed FoVs. Binned medians and fitted log-space trends show an overall decrease of integrated excess power with increasing canopy filling factor in both diagnostics.}
\label{fig:supp_int_excess_ff_obs}
\end{figure*}

The observational results (Fig.~\ref{fig:supp_int_excess_ff_obs}) show the same overall behaviour. For each diagnostic, 110 valid windows are obtained in both FoVs. The quiet-Sun windows are concentrated at low canopy filling factor, with median \(ff \approx 0.028\) and a range of 0.00--0.20, whereas the active-region windows occupy a predominantly canopy-filled regime, with median \(ff \approx 0.91\) and a range of 0.37--1.00. In Ca~{\sc ii}~8542\,\AA, the median integrated excess is \(\approx 4.01\) in the quiet Sun and \(\approx 0.54\) in the active region, corresponding to a quiet-Sun to active-region ratio of \(\approx7.5\) (0.87\,dex). In H$\alpha$, the corresponding medians are \(\approx 3.49\) and \(\approx 0.44\), respectively, corresponding to a factor of \(\approx8.0\) (0.90\,dex). The fitted log-space trends are negative for both regions and both diagnostics, again showing an overall decrease of 3--5\,mHz excess power with increasing canopy filling factor.

Although the observations and simulations show the same overall $EP$--$ff$ behaviour, their absolute $EP$ amplitudes are not expected to agree one-to-one. The largest difference occurs for Ca~{\sc ii}~8542\,\AA\ in the low-$ff$ regime, where the quiet-Sun observational excess is approximately 3--4 times larger than in the canopy-poor simulation. This discrepancy is not similarly present in H$\alpha$, and is much smaller for Ca~{\sc ii}~8542\,\AA\ in the canopy-rich regime, arguing against a generic normalisation or data-reduction offset. First, $EP$ measures the enhancement relative to a fitted spectral background through $P/P_{\rm fit}$ rather than the absolute oscillatory power, so differences in both the spectral background and the enhancement itself can affect its amplitude. Second, quiet-Sun Ca~{\sc ii}~8542\,\AA\ is particularly responsive to intermittent acoustic-shock dynamics, which can produce substantial temperature and velocity perturbations and corresponding changes in its inner-wing intensity \citep[e.g.][]{2022A&A...668A.153M}. At the fixed $-200$\,m\AA\ sampling used here, thermodynamic and opacity variations together with Doppler-induced profile shifts can therefore enhance the measured intensity fluctuations. H$\alpha$ has a different, strongly scattering-dominated chromospheric response, with its opacity more closely coupled to density than to local temperature \citep{2012ApJ...749..136L}, so the same perturbations need not produce an equally strong relative intensity response. Finally, the observed quiet Sun and the canopy-poor Bifrost atmosphere are not one-to-one realisations; differences in shock occurrence, thermodynamic stratification, spectral-background shape, and detailed line formation can therefore affect the absolute $EP$ values. We consequently interpret Figs.~\ref{fig:supp_int_excess_ff} and \ref{fig:supp_int_excess_ff_obs} primarily in terms of their common topology-dependent $EP$--$ff$ behaviour rather than requiring exact agreement of their absolute $EP$ amplitudes.

The observational distributions show greater scatter than their simulation counterparts. This is expected given the presence of observational noise, unresolved fine-scale structure, and genuine spatial and temporal variability, together with uncertainties introduced by the lower spatial resolution of the HMI-based magnetic information and by the extrapolation itself at chromospheric heights. We therefore regard the observational filling factors as topology proxies and focus on the overall $EP$--$ff$ relationship rather than on detailed agreement between individual windows or fitted slopes.

Overall, Figs.~\ref{fig:supp_int_excess_ff} and \ref{fig:supp_int_excess_ff_obs} provide compact quantitative support for the central claim in both simulations and observations: the measured large-scale three-minute power in chromospheric intensity diagnostics is strongly modulated by canopy areal filling, with canopy-poor scenes occupying the high-excess, low-\(ff\) regime and canopy-dominated scenes occupying the low-excess, high-\(ff\) regime.

\section{Discussion and conclusions}

The measured FoV-averaged (large-scale) three-minute power in the chromosphere is not set by subsurface driving alone; it is strongly modulated by the overlying magnetic topology -- the umbrella effect. Across high-resolution observations and realistic simulations, we find a consistent pattern: extended quiet-Sun scenes (magnetically quiet over the FoV and its immediate environment, implying minimal influence from externally rooted canopies; cf.~\citealt{2021RSPTA.37900174J}) show pronounced $\sim$5\,mHz enhancements, whereas canopy-dominated scenes do not. This topological control is evident across diagnostics (H$\alpha$, Ca\,\textsc{ii}\,8542\,\AA, synthetic observables) and physical measures (intensity, temperature, LOS velocity).

In quiet-Sun observations, FoV-averaged power peaks at 4.5--5\,mHz, most clearly at mid-chromospheric heights sampled by intermediate wing positions in intensity, with a coherent height dependence reproduced in bisector-derived LOS velocities (peak near 20--30\% levels). The coronal-hole simulation echoes this behaviour: height-resolved LOS velocity and temperature spectra show a chromospheric maximum near 5\,mHz, with photospheric power peaks at lower frequencies that shift upward with height before declining towards the transition region. Notably, FoV averages were computed by averaging pixel-wise power spectra (not time series before spectral analysis), so the presence or absence of a 3--5\,mHz peak is not an artefact of phase cancellation.

By contrast, in the active-region observations, large-scale 3--5\,mHz enhancements are absent in both intensity and bisector-velocity spectra. In the enhanced-network simulation, three-minute peaks are likewise absent in synthetic intensities and in height-sampled LOS-velocity and temperature spectra. Side-on field extrapolations (observations) and simulated field renderings both show dense, multi-height canopies with high lateral continuity over the analysis region. 

Our results indicate that a key parameter controlling the measured FoV-averaged three-minute signature is the canopy's lateral continuity, or areal filling, at chromospheric heights. Dense, inclined canopies reduce the measured FoV-averaged 3--5\,mHz power, likely through a combination of altered propagation conditions, refraction or partial reflection, mode conversion, and phase mixing, whereas porous or vertically intermittent canopies allow a pronounced 3--5\,mHz enhancement to remain measurable. We emphasise that this FoV-averaged canopy-modulation effect concerns canopy-dominated regions outside compact strong-field cores and does not contradict the well-established presence of strong local three-minute oscillations in sunspot umbrae or pores with predominantly vertical fields. These results extend the concept of magnetic shadows \citep{2001ApJ...561..420M} from localised suppression to a FoV-averaged, topology-governed reduction of the measured three-minute signature that can be imposed by structures originating beyond the nominal FoV.

They also reconcile ALMA reports of strong three-minute power only in magnetically quiet fields \citep{2021RSPTA.37900174J}: formation-height variations alone are insufficient to explain the full behaviour, while magnetic topology, particularly the presence or absence and areal filling of overlying canopies, provides a natural control parameter (see Fig.~\ref{fig:Supp_alma_sim}). The reduction of the observable 3--5\,mHz signature is therefore not tied to a specific wavelength or diagnostic.

To avoid over-interpreting these results, we emphasise that the umbrella effect describes a topology-dependent modulation of the measured FoV-averaged three-minute power and does not uniquely identify a single microphysical mechanism. Several processes may contribute, including refraction and partial reflection as waves encounter regions where the canopy brings the plasma-$\beta$ towards unity, mode conversion into components that are less strongly expressed in the analysed compressive diagnostics, and phase mixing in laterally extended, inclined fields. In addition, canopy inclination modifies the effective acoustic cutoff frequency through the ramp effect, altering which periods can propagate upward along field-guided paths. Distinguishing the relative importance of these processes is a natural next step; here we establish the net observable consequence that laterally continuous, multi-height canopies substantially reduce the measured FoV-averaged three-minute signature across diagnostics and in both observations and simulations.

We also note that photospheric $p$-mode power is known to be suppressed locally within concentrated magnetic features (e.g. magnetic bright points and pores; \citealt{2012ApJ...744...98C, 2014ApJ...796...72J, 2016ApJ...823...45K}). Such local suppression can contribute to FoV-averaged power only insofar as these features occupy a substantial area fraction. Our key result, however, concerns the measured chromospheric FoV-averaged three-minute power and its dependence on canopy topology across multiple heights and diagnostics, supported independently by the canopy-rich versus canopy-poor simulation comparison.

A key implication is that chromospheric three-minute power is a topology-weighted observable rather than a direct proxy for upward wave-energy flux: dense, laterally continuous canopies can substantially reduce the measured FoV-averaged three-minute power even when waves are present locally. Quantitative energy-flux estimates require additional assumptions (e.g. mode identification and response functions) and are therefore beyond the scope of the present Article; here we establish the canopy-controlled modulation of the measured three-minute signature and a practical criterion for when large-scale three-minute power should, or should not, be measurable. This complements studies that quantify chromospheric wave-energy transport and dissipation as a function of local magnetic-field inclination and strength \citep[e.g.][]{2024ApJ...965..136K}: those studies address the energetics of locally measured waves, whereas the present work addresses how canopy areal filling and lateral continuity modulate the measured FoV-averaged three-minute power before any inference of wave-energy flux is made.

Finally, the framework is predictive and testable: for otherwise comparable atmospheric and observational conditions, regions where canopies achieve high areal filling at chromospheric heights are expected to show reduced FoV-averaged 3--5\,mHz power, whereas regions with sparse or vertically intermittent canopies should more readily retain a pronounced $\sim$5\,mHz enhancement. This criterion is supported quantitatively by both the simulations and observations (Figs.~\ref{fig:supp_int_excess_ff} and \ref{fig:supp_int_excess_ff_obs}): in each case, the integrated 3--5\,mHz excess shows an overall decrease with increasing canopy filling factor. The present observational comparison is not a statistical survey of all quiet-Sun and active-region conditions, and a broader survey across a wider range of canopy morphologies will be required to quantify the dependence on canopy filling factor, field inclination, magnetic-field strength, and the presence of compact strong-field cores. A detailed phase and coherence analysis across multiple chromospheric diagnostics (accounting for their response functions and formation-height ranges) is also an important next step for assessing upward propagation and wave-mode partition in canopy-dominated scenes. More generally, this umbrella effect applies to any vertical coupling between lower and upper layers: the measured oscillatory signature can be reduced at any height where canopy areal filling becomes high, as set by footpoint field strengths and the 3D topology. High-resolution, multi-height spectroscopy from \textsc{Sunrise~iii} \citep{2025SoPh..300...75K, Solanki2026} and the Daniel K. Inouye Solar Telescope (DKIST; \citealt{2021SoPh..296...70R}) can directly test this criterion by co-mapping canopy inclination and areal coverage with FoV-averaged power. An analogous topology-governed transfer function may operate in magnetised cool stars with extended atmospheric canopies, and should be considered when helio- or asteroseismic analyses relate interior driving to chromospheric observables.

\begin{acknowledgements}
SJ and DBJ wish to thank the UK Science and Technology Facilities Council (STFC) for
the consolidated grants ST/T00021X/1 and ST/X000923/1. 
DBJ also acknowledges support from the Leverhulme Trust via the Research Project Grant RPG-2019-371, and from the UK Space Agency via the National Space Technology Programme (grant SSc-009). 
LRvdV, SW, TMDP, MS, and ERU acknowledge support from the Research Council of Norway through its Centres of Excellence scheme, project number 262622, and acknowledge the computational resources provided by UNINETT Sigma2 - the National Infrastructure for High Performance Computing and Data Storage in Norway.
We are grateful to Theodore (Ted) Tarbell, Donald Schmit, and Peter Sütterlin for acquiring the SST observations used in this study. We particularly acknowledge Ted Tarbell, whose efforts were instrumental in making these observations possible. Ted passed away in 2019, and we dedicate this work to his memory.
We wish to acknowledge scientific discussions with the Waves in the Lower Solar Atmosphere (WaLSA; \href{https://WaLSA.team}{www.WaLSA.team}) team, which has been supported by the Research Council of Norway (project no. 262622), The Royal Society (award no. Hooke18b/SCTM; \citealt{2021RSPTA.37900169J}), and the International Space Science Institute (ISSI Team 502).
\end{acknowledgements}

\bibliographystyle{aa}
\bibliography{aa61881-26}

\begin{appendix}

\section{Spatial maps of 3--5\,mHz power}

\begin{figure*}[!hb]
\centering
\includegraphics[width=1.0\textwidth]{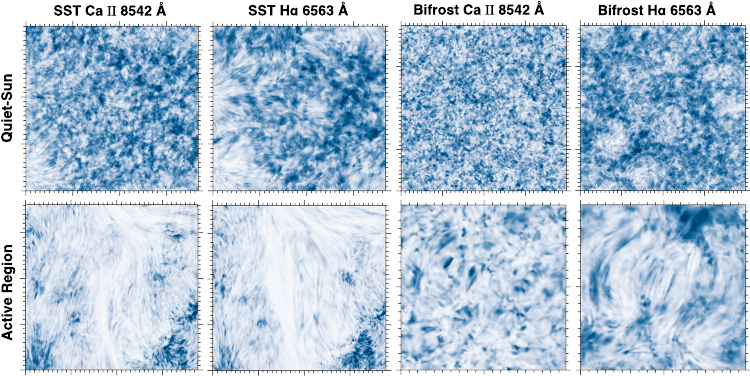}
\caption{Representative spatial maps of three-minute band power in observations and simulations.
Pixel-wise maps show band-integrated three-minute power, obtained by integrating the FFT power spectral density over 3--5\,mHz (a 2\,mHz-wide window centred on the three-minute band). The figure is arranged in two rows (top: quiet Sun and coronal hole; bottom: active region and enhanced network) and four columns (left to right): SST Ca~{\sc ii}~8542\,\AA\ intensity at $-200$\,m\AA, SST H$\alpha$ intensity at $-400$\,m\AA, Bifrost synthetic Ca~{\sc ii}~8542\,\AA\ intensity at the corresponding offset, and Bifrost synthetic H$\alpha$ intensity at the corresponding offset. The SST FoV is 38$^{\prime\prime}\times$36$^{\prime\prime}$. The Bifrost coronal-hole maps show the full 24$\times$24\,Mm domain, whereas the enhanced-network maps show the central 12$\times$12\,Mm region (as marked in Fig.~\ref{fig:AR_sim}) to exclude the quieter surrounding areas. The blue--white colour scale indicates increasing-to-decreasing band power (blue: strong; white: weak or absent).}
\label{fig:Supp_bandpower}
\end{figure*}

To visualise how the three-minute signal is distributed spatially within each scene, we computed pixel-wise maps of 3--5\,mHz band-integrated power by integrating the FFT power spectral density over 3--5\,mHz (a 2\,mHz-wide window centred on the three-minute band). Because producing maps for all diagnostics and wavelength positions analysed in the main text would be impractically large, we show a compact representative set for the chromospheric sampling positions that exhibit the strongest three-minute enhancements in the quiet-Sun case: Ca~{\sc ii}~8542\,\AA\ at $-200$\,m\AA\ and H$\alpha$ at $-400$\,m\AA. Fig.~\ref{fig:Supp_bandpower} compares these maps between the quiet, canopy-poor and the active, canopy-dominated environments analysed here for both SST observations and Bifrost simulations. While spatial variability exists within each FoV, the quiet, canopy-poor scenes show broadly distributed three-minute power, whereas canopy-dominated scenes show markedly reduced band power over most pixels, consistent with a canopy-controlled modulation of the measured FoV-averaged three-minute signature.

\section{Synthetic ALMA diagnostics from simulations}
\label{sec:alma_synthetic}
To complement the main analysis, we present synthetic diagnostics at millimetre wavelengths corresponding to ALMA Band 6 (1.2\,mm) and Band 3 (3.0\,mm), derived from the same Bifrost simulations described in the main text.

\begin{figure*}[h]
\centering
\includegraphics[width=1.0\textwidth]{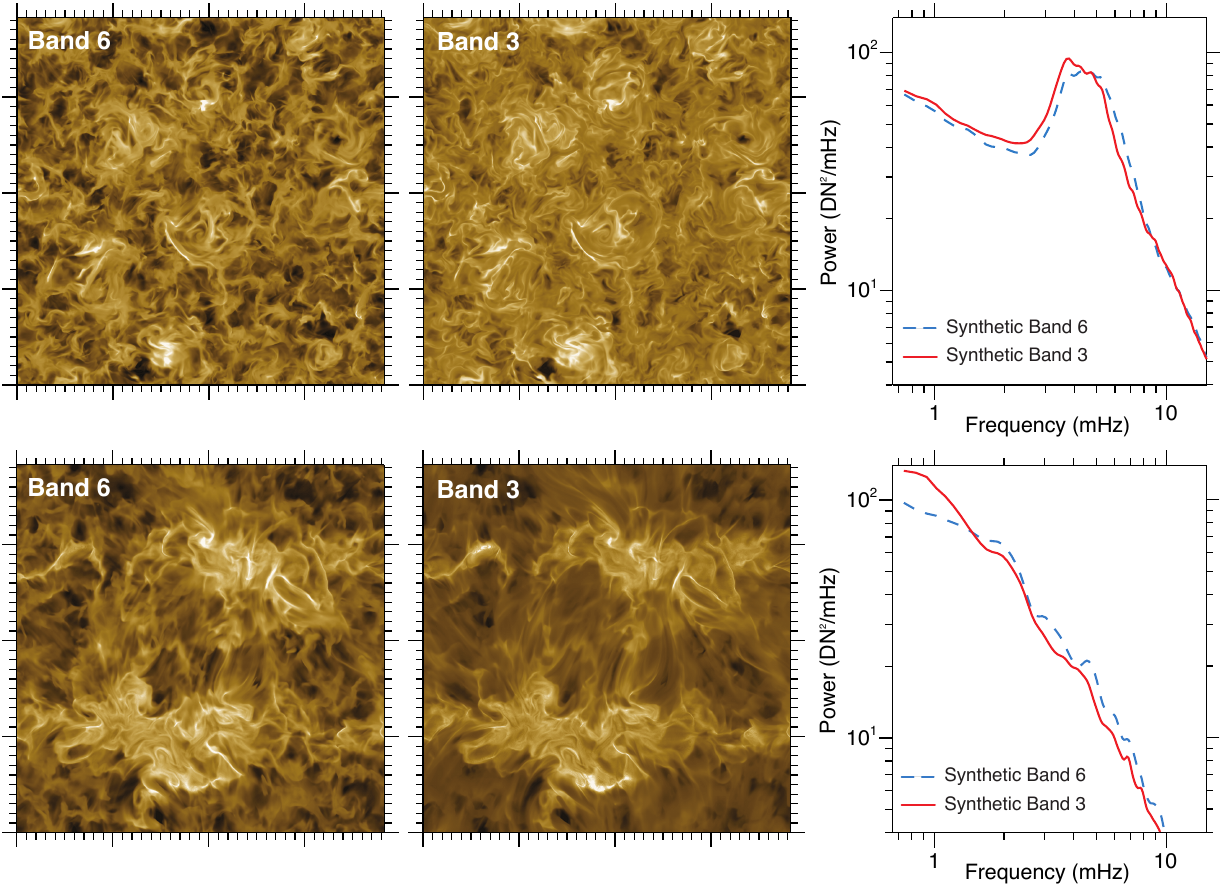}
\caption{Synthetic ALMA diagnostics from Bifrost simulations.
Top row: Simulated brightness temperature maps at ALMA Band~6 (1.2\,mm; left) and Band~3 (3.0\,mm; middle) from the coronal-hole simulation, alongside their spatially averaged power spectra (right). Both bands exhibit pronounced oscillatory power enhancements peaking between 4--5\,mHz, consistent with large-scale three-minute signatures in magnetically quiet chromospheric conditions.
Bottom row: Corresponding synthetic ALMA diagnostics from the enhanced-network simulation. No significant oscillatory power enhancement is detected at either band, highlighting the reduction of the measured large-scale oscillatory signature in the canopy-rich atmosphere.}
\label{fig:Supp_alma_sim}
\end{figure*}

Fig.~\ref{fig:Supp_alma_sim} summarises the results. The top row shows synthetic brightness temperature maps from the coronal-hole simulation at Band 6 and Band 3 (left and middle), along with their FoV-averaged power spectra (right). Both ALMA bands exhibit clear power enhancements in the 4--5\,mHz band, consistent with the presence of large-scale three-minute oscillations in the mid-to-upper chromosphere.

By contrast, the bottom row displays the corresponding results from the enhanced-network simulation. In this magnetically complex scenario, no significant power enhancement is visible at either ALMA band, indicating suppression of large-scale oscillations due to the presence of extended chromospheric canopy fields. As in the main text, we restricted the analysis to the core magnetised region (see Fig.~\ref{fig:AR_sim} of the main article), explicitly excluding surrounding quiet areas.

These results reinforce the findings of \citet{2021RSPTA.37900174J}, while now providing a numerical foundation for the observed variability of three-minute power across ALMA fields of view. Together with the multi-height optical diagnostics in the main paper, these millimetre-band results further support the conclusion that magnetic topology -- rather than formation height alone -- modulates the measured FoV-averaged chromospheric oscillation power. We note that millimetre continua have broad contribution functions; the topology-driven differences reported here remain even under that broad sampling, in agreement with the height-resolved optical diagnostics.

\section{Spectral contribution functions and formation heights}
\label{sec:appendix_formation_height}
\begin{figure*}[h]
\centering
\includegraphics[width=0.94\textwidth]{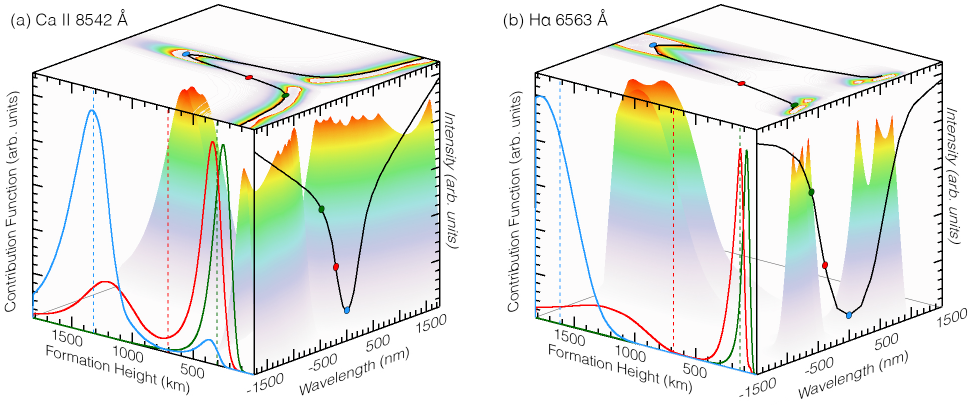}
\caption{Contribution-function analysis of Ca\,\textsc{ii}\,8542\,\AA\ (a) and H$\alpha$\,6563\,\AA\ (b) in the FALC atmosphere. Observed profiles and selected wavelength positions are shown on the right; the corresponding RH non-LTE line-depression contribution functions are shown in the cubes and as functions of height on the left. The top panels give the mean formation height across each profile. Dashed vertical lines on the left panel mark the mean formation heights of the selected wavelengths.}
\label{fig:Supp_CF_cube}
\end{figure*}

\begin{figure*}[th!]
\centering
\includegraphics[width=0.85\textwidth]{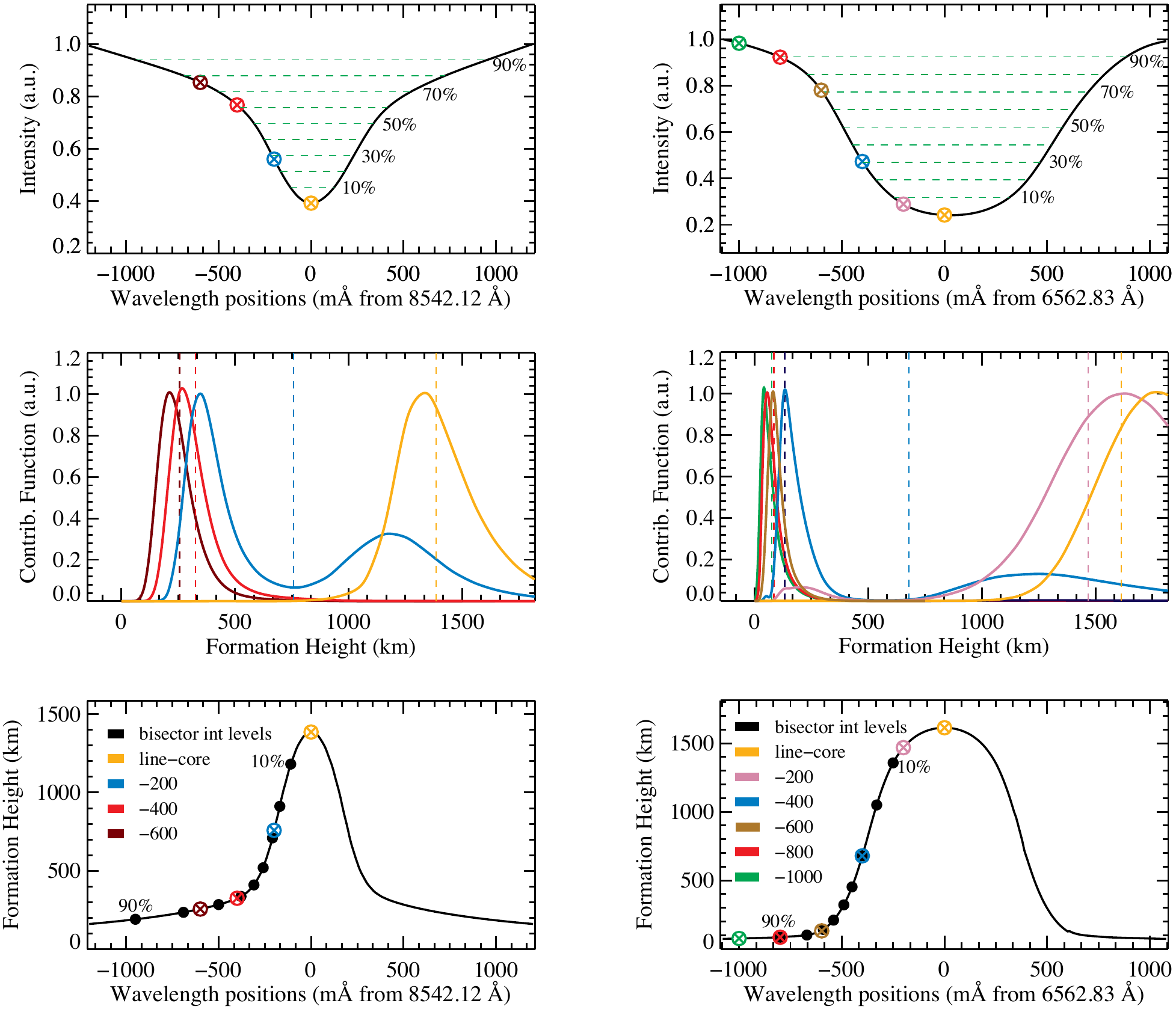}
\caption{Contribution functions and formation heights for the quiet-Sun intensity and bisector diagnostics. Top: spatially averaged Ca\,\textsc{ii}\,8542\,\AA\ and H$\alpha$ profiles with the selected wavelength positions and bisector intensity levels. Middle: corresponding line-depression contribution functions computed in FALC and convolved with the CRISP transmission profiles; dashed lines mark the mean formation heights. Bottom: mean formation height versus wavelength for the fixed-wavelength and bisector diagnostics. Bisectors are shown on the averaged profiles only for illustration; the velocity analysis is performed independently at each spatial pixel.}
\label{fig:Supp_CFs}
\end{figure*}

To support the interpretation of height-dependent oscillatory power across both intensity and Doppler diagnostics, Fig.~\ref{fig:Supp_CF_cube} provides a compact 3D visualisation of the CF computation and representative formation heights, while Fig.~\ref{fig:Supp_CFs} presents the full set of CFs and formation-height curves for all sampled positions used in the analysis.

Fig.~\ref{fig:Supp_CF_cube} summarises the CF workflow in FALC (with CRISP transmission applied) for Ca\,\textsc{ii}\,8542\,\AA\ and H$\alpha$. Each cube links the observed line profile (right) to the 3D CF surfaces (interior) and the mean formation-height curves (top) for three representative wavelength samples (coloured markers). The CF surfaces highlight vertical extent and peak location; the left panels show the corresponding CFs collapsed versus height for direct comparison. This schematic is a guide to our sampling strategy.

In Fig.~\ref{fig:Supp_CFs}, the top panels show the spatially averaged spectra for the quiet-Sun observations in Ca\,\textsc{ii}\,8542\,\AA\ (left) and H$\alpha$ (right), with the selected wavelength positions marked in colour. The fractional intensity levels used for bisector analysis (10\% to 90\% of the line depth) are also overlaid as horizontal dashed lines, mapped onto the corresponding profile depths.

The middle panels display the line-depression CFs for each selected wavelength position, computed using the RH radiative transfer code within the FALC model atmosphere, after convolution with the CRISP instrumental transmission profiles used in the observations. These CFs illustrate the vertical extent and peak contribution of each spectral sample, with vertical dashed lines marking the average formation height at each position.

The bottom panels show the mean formation heights for all sampled points across the spectral profile. Markers highlight both the selected fixed-wavelength positions and the bisector intensity levels, illustrating their approximate correspondence to different chromospheric heights. While the CFs exhibit broad vertical extents, the indicated mean values provide a useful height proxy for connecting observed oscillatory signatures to their representative formation layers.

Due to the broad cores of the H$\alpha$ and Ca\,\textsc{ii}\,8542\,\AA\ lines, LOS velocities are not derived directly at the line centre, where spectral reversals and core asymmetries may hinder reliable bisector estimation. As a result, the 10\% intensity level used in the bisector analysis is formed slightly below the nominal core height. Nevertheless, as shown in the bottom panels, the bisector-derived velocities at 10\%--40\% intensity levels generally correspond to heights spanning from the mid-to-upper chromosphere down to the lower chromosphere, respectively, and should be interpreted as height proxies rather than fixed geometric layers at every pixel.

\end{appendix}

\end{document}